\documentclass{aa}
\usepackage{epsfig,times}

\makeatletter
\def\@cite#1#2{(#1\if@tempswa , #2\fi)}
\def\@citex[#1]#2{\if@filesw\immediate\write\@auxout{\string\citation{#2}}\fi
  \def\@citea{}\@cite{\@for\@citeb:=#2\do
    {\@citea\def\@citea{;\penalty\@m\ }\@ifundefined
       {b@\@citeb}{{\bf ?}\@warning
       {Citation `\@citeb' on page \thepage \space undefined}}%
\hbox{\csname b@\@citeb\endcsname}}}{#1}}
\makeatother

\title{Unusually rapid variability of the GRB000301C optical afterglow
\thanks{Based on observations collected at the Bologna Astronomical
Observatory in Loiano, Italy and at the TNG, Canary Islands, Spain}}

\author{N. Masetti\inst{1}
\and C. Bartolini\inst{2}
\and S. Bernabei\inst{3}
\and A. Guarnieri\inst{2}
\and E. Palazzi\inst{1}
\and E. Pian\inst{1}
\and A. Piccioni\inst{2}
\and A.J. Castro-Tirado\inst{4,5}
\and J.M. Castro Cer\'on\inst{6}
\and L. Verdes-Montenegro\inst{4}
\and R. Sagar\inst{7} 
\and V. Mohan\inst{7}
\and A.K. Pandey\inst{7}
\and S.B. Pandey\inst{7}
\and H. Bock\inst{8}
\and J. Greiner\inst{9}
\and S. Benetti\inst{10}
\and R.A.M.J. Wijers\inst{11} 
\and G.M. Beskin\inst{12} 
\and J. Gorosabel\inst{13}}

\institute{Istituto Tecnologie e Studio Radiazioni Extraterrestri, CNR,
Via Gobetti 101, I-40129 Bologna, Italy
\and
Dipartimento di Astronomia, Universit\'a di Bologna, Via Ranzani 1,
I-40127 Bologna, Italy
\and
Osservatorio Astronomico di Bologna, Via Ranzani 1, I-40127
Bologna, Italy
\and
IAA-CSIC, P.O. Box 03004, E-18180, Granada, Spain 
\and 
LAEFF-INTA, Madrid, Spain
\and
Real Instituto y Observatorio de la Armada, 11110 San Fernando Naval,
C\'adiz, Spain
\and
Uttar Pradesh State Observatory, Manora Peak, Nainital, 263 129, India
\and
Landessternwarte Heidelberg, Heidelberg, Germany
\and
Astrophysikalisches Institut, 14482 Potsdam, Germany
\and
TNG Observatory, Canary Islands, Spain
\and
Dept. of Physics \& Astronomy, SUNY, Stony Brook, NY 11794-3800, USA
\and
Special Astrophysical Observatory of RAS, Nizhnij Arkhyz,
Karachai--Cherkessia, 357147 Russia
\and
Danish Space Research Institute, Copenhagen, Denmark
}

\offprints{Nicola Masetti, masetti@tesre.bo.cnr.it}
\date{Received April 13, 2000; Accepted June 26, 2000}
\thesaurus{03(13.07.1)}

\begin{document}

\maketitle
\markboth{N. Masetti et al.: Unusually rapid variability of the 
GRB000301C optical afterglow}{}

\begin{abstract}

We present $BVRI$ light curves of the afterglow of GRB000301C, one of the
brightest ever detected at a day time scale interval after GRB trigger.
The monitoring started 1.5 days after the GRB and ended one month later. 
Inspection of the extremely well sampled $R$ band light curve and
comparison with $BVI$ data has revealed complex behavior, with a long
term flux decrease and various short time scale features superimposed.
These features are uncommon among other observed afterglows, and might
trace either intrinsic variability within the relativistic shock
(re-acceleration and re-energization) or inhomogeneities in the medium in
which the shock propagates.


\end{abstract}

\section{Introduction}

Fundamental progress on the knowledge of Gamma-Ray Bursts (GRBs) has been
made possible by detection of their optical counterparts.  Of nearly 40
GRBs accurately and rapidly localized so far by BSAX, BATSE/RXTE, IPN,
and promptly followed up in the optical, only about 50\% exhibited optical
afterglows\footnote{http://www.aip.de/$\sim$jcg/grbgen.html}, suggesting
that these sources are rapidly fading, or heavily obscured. 
The best monitored afterglows (GRBs 970228, 970508, 980326, 980519,
990123, 990510) exhibit 
a variety of behaviors, indicating that the shape of the optical
decay must be determined not only by the intrinsic physics, but also by
the nature, structure and composition of the surrounding medium. 
Therefore, optical light curves of GRB counterparts need to be frequently
sampled for long time intervals, to follow the evolution of the afterglow
and to allow mapping the characteristics of the medium.

GRB000301C was detected by the IPN and by the RXTE ASM on 2000 March 1.4
UT with an error box of 50 arcmin$^2$ (Smith et al. 2000). Its field was
acquired starting $\sim$1.5 days later by various optical, infrared and
radio telescopes.  The optical afterglow was independently detected by 
Fynbo et al. (2000) and by us (Bernabei et al. 2000a), and is among the
brightest ever observed. Near-infrared detection and monitoring of the
afterglow are reported in Rhoads \& Fruchter (2000). Observations of the
counterpart at radio and millimetric wavelengths have been reported by
Berger \& Frail (2000) and Bertoldi (2000), respectively.  Ultraviolet
spectroscopy with the STIS instrument onboard HST allowed the
determination of the redshift (Smette et al. 2000), then refined by
optical ground-based spectroscopy ($z$ = 2.03, Castro et al. 2000).
The good sampling and the brightness of the GRB000301C afterglow have
allowed a detailed study of its evolution up to 15 days after the
explosion. In this paper we present the results of the optical monitoring
conducted at Loiano, Calar Alto, Sierra Nevada, Nainital and Canary
Islands.

\begin{table*}
\caption[]{Journal of the optical observations of the GRB000301C 
afterglow}
\begin{center}
\begin{tabular}{rccccc}
\noalign{\smallskip}
\hline
\noalign{\smallskip}
Exposure start & Telescope & Filter & Exp. time &
Seeing & Magnitude$^1$
\\
\multicolumn{1}{c}{(UT)} & & & (minutes) & (arcsecs) &  \\
\noalign{\smallskip}
\hline
\noalign{\smallskip}

2000 Mar 2.906 & UPSO     & R & 70 & 1.4 & 20.42 $\pm$ 0.04$^2$ \\
3.144 & CAHA   & R & 5    & 1.1 & 20.25 $\pm$ 0.05 \\
3.179 & CAHA   & B & 15   & 1.1 & 21.07 $\pm$ 0.05 \\
3.185 & Loiano & R & 16.7 & 2   & 20.16 $\pm$ 0.05 \\
3.205 & CAHA   & R & 5    & 1.1 & 20.25 $\pm$ 0.05 \\
3.210 & CAHA   & I & 10   & 1.1 & 19.94 $\pm$ 0.07 \\
3.219 & CAHA   & V & 15   & 1.1 & 20.57 $\pm$ 0.05 \\
3.232 & CAHA   & B & 15   & 1.1 & 21.10 $\pm$ 0.12 \\
3.913 & UPSO   & R & 50   & 1.2 & 20.51 $\pm$ 0.04 \\
4.038 & CAHA   & R & 15   & 1.6 & 20.53 $\pm$ 0.06 \\
4.149 & Loiano & R & 36.7 & 3   & $>$ 20.25$^3$ \\
4.165 & Loiano & B & 20   & 3   & $>$ 21.0  \\
5.135 & SNO    & R & 20   & 2   & 20.47 $\pm$ 0.07 \\
5.152 & SNO    & B & 20   & 2   & 21.60 $\pm$ 0.20 \\
5.172 & SNO    & V & 20   & 2   & 21.04 $\pm$ 0.20 \\
5.930 & UPSO   & R & 85   & 1.3 & 21.14 $\pm$ 0.06 \\
6.135 & Loiano & R & 30   & 1.7 & 21.65 $\pm$ 0.20 \\
6.163 & Loiano & B & 30   & 1.7 & 22.45 $\pm$ 0.15 \\
6.185 & Loiano & I & 16.7 & 1.7 & 20.82 $\pm$ 0.15 \\  
6.968 & UPSO   & R & 35   & 1.6 & $>$ 21.6 \\
7.125 & Loiano & R & 30   & 1.7 & 21.68 $\pm$ 0.15 \\
7.149 & Loiano & B & 35   & 1.7 & 22.43 $\pm$ 0.10 \\
7.177 & Loiano & I & 20   & 1.7 & 21.20 $\pm$ 0.15 \\
7.894 & UPSO   & R & 105  & 1.6 & 22.00 $\pm$ 0.15 \\
8.146 & Loiano & R & 30   & 1.6 & 21.68 $\pm$ 0.10 \\
8.170 & Loiano & I & 30   & 1.6 & 21.61 $\pm$ 0.10 \\
8.924 & UPSO   & R & 75   & 1.3 & 22.04 $\pm$ 0.20 \\
Apr 5.213 & TNG & B & 20  & 0.5 & $>$ 25.5 \\
\noalign{\smallskip}
\hline
\noalign{\smallskip}
\multicolumn{6}{l}{$^1$Magnitudes of the GRB counterpart, not corrected
for interstellar absorption}\\
\multicolumn{6}{l}{$^2$Uncertainties of the magnitudes are at
1$\sigma$ confidence level; lower limits at 3$\sigma$}\\
\multicolumn{6}{l}{$^3$Note that this measurement is reported as a
detection in Bernabei et al. (2000b)}\\

\end{tabular}
\end{center}
\end{table*}

\begin{figure*}
\vspace{-1.5cm}
\begin{center}
\epsfig{figure=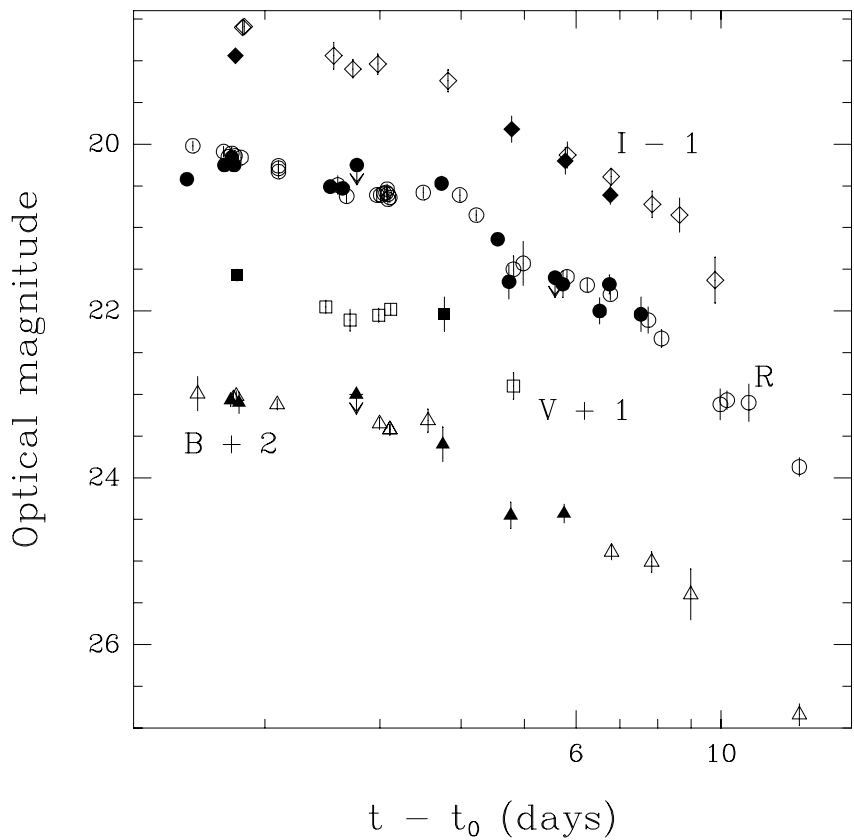,width=14cm}
\end{center}
\vspace{-1.8cm}
\caption[]{$BVRI$ light curves of GRB000301C afterglow, based on the data
presented in this paper and in the literature (see text). 
Filled symbols represent data presented in this work, while open symbols
refer to measurements published by other authors. We have
consistently referred all magnitudes to the calibration zero point of
Henden (2000). To the statistical uncertainties a 5\% systematic error has
been added in quadrature (see text). No Galactic extinction correction,
nor host galaxy flux subtraction has been applied. The GRB start time,
indicated with $t_0$, corresponds to 2000 March 1.410845 UT}
\end{figure*}

\begin{figure}
\vspace{-1.2cm}
\begin{center}
\epsfig{figure=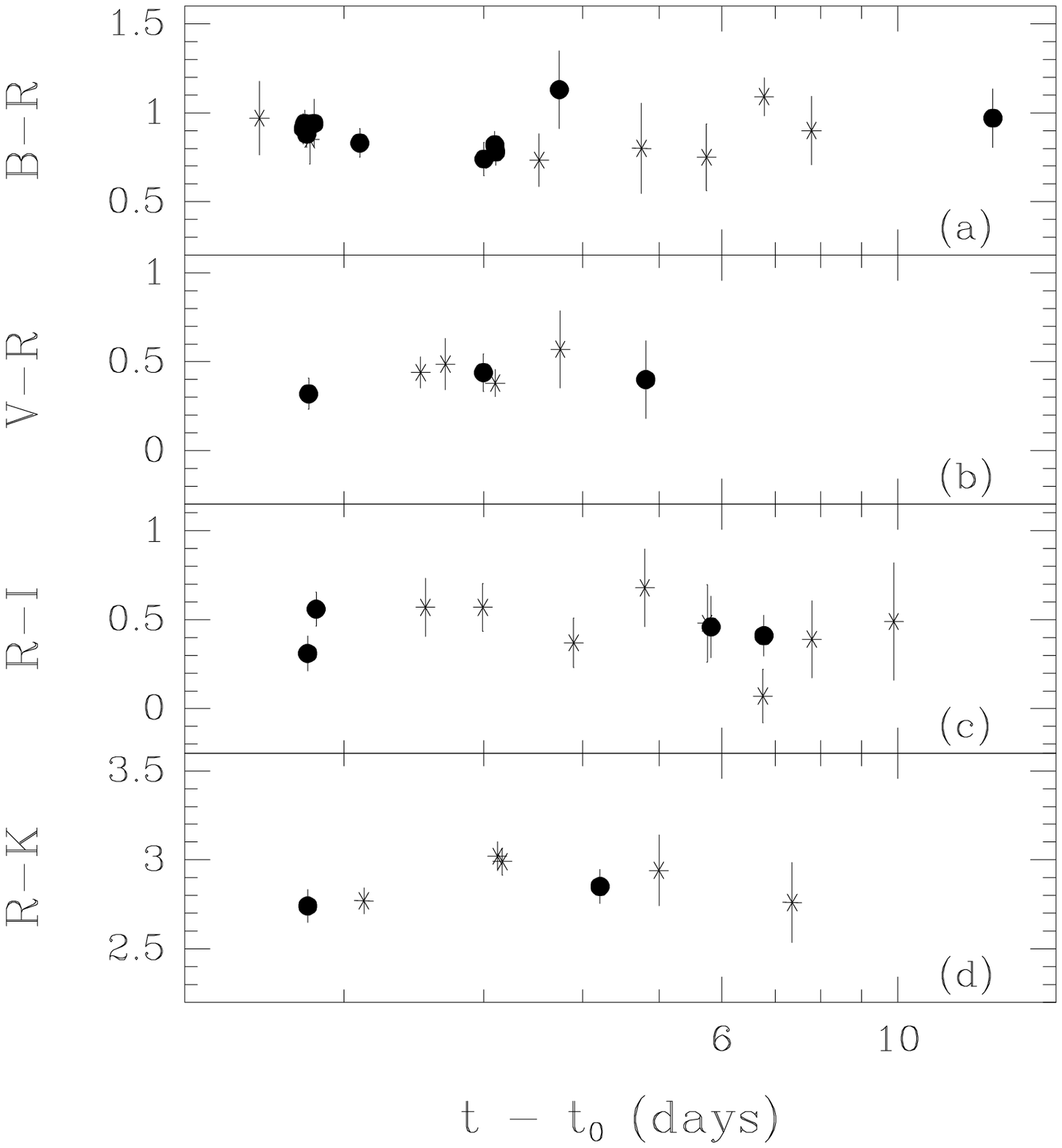,width=11.5cm}
\end{center}
\vspace{-1.8cm}
\caption[]{Colors of GRB000301C afterglow (data are from this paper and 
from the literature, see text).  These are reported as filled
circles when computed between pairs of measurements spaced apart in time
by no more than 0.5 hr, and as stars when the temporal separation
is larger than 0.5 hr, but smaller than 9 hr. As in Fig. 1, calibration by
Henden (2000) has been adopted and a 5\% systematic error has been added
in quadrature (see text).  The GRB start time, indicated with $t_0$,
corresponds to 2000 March 1.410845 UT}

\end{figure}

\section{Observations and data reduction}

Optical $BVRI$ images were collected soon after the notification of the
GRB000301C detection, starting $\sim$1.5 days after the high-energy event.
Observations were carried out with CCD cameras at the 1.52-meter 
``G.D. Cassini" telescope of the Bologna University in
Loiano, Italy, at the 1.0-meter UPSO telescope in Nainital, India, 
at the 1.2-meter CAHA telescope in Calar Alto, Spain, at the
1.5-meter telescope of the Sierra Nevada Observatory (SNO) in Granada,
Spain, and at the {\it Telescopio Nazionale Galileo} (TNG) in the Canary
Islands, Spain.
The complete log of these observations is reported in Table 1.

Images were debiased and flat-fielded with the standard cleaning
procedure.
Due to the proximity (7 arcsec) of the target to a bright star, we chose
to use standard PSF-fitting as our photometric technique, and for this
purpose we used the DAOPHOT II image data analysis package PSF-fitting
algorithm (Stetson 1987) running within {\sl MIDAS}.
A two-dimensional gaussian with two free parameters (the half width at
half maxima along $x$ and $y$ coordinates of each frame) was modeled on at
least 5 non-saturated bright stars in each image. For each filter, this
procedure yields frames magnitude differences among the photometric
references of less than 1\% in all frames.
The errors associated with the measurements reported in Table 1 represent 
statistical uncertainties (at 1$\sigma$), obtained with the
standard PSF-fitting procedure. 
Calibration was done using the $BVRI$ magnitudes of field stars as
measured by Henden (2000).

\section{Results}

In our images we detect a point-like source within the 50 square arcmin
error box of the GRB, at the position RA= 16$^{\rm h}$ 20$^{\rm m}$
18$\fs$5, Dec = 29$^{\circ}$ 26$'$ 35$''$ (J2000), consistent with that
given by Fynbo et al. (2000).
The source variability (see magnitude levels in Table 1) suggests that
this is the afterglow of GRB000301C. The light curves in $BVRI$
bands are reported in Fig. 1, where our data are complemented with those
published by other authors (Sagar et al. 2000, Jensen et al. 2000, and the
GCN circulars 
archive\footnote{http://gcn.gsfc.nasa.gov/gcn/gcn3\_archive.html}).
Note that some results presented in this paper supersede preliminary
values reported in GCNs. Our analysis of UPSO $R$ band data yielded
results consistent with those reported by Sagar et al. (2000). No
correction has been applied for Galactic extinction, which is anyway small
in the direction of the GRB ($E(B-V)$ = 0.052, Schlegel et al. 1998); nor
has been subtracted any host galaxy continuum emission, this being
negligible (Fruchter et al. 2000).
A 5\% systematic error was added in quadrature to the errors reported in
Figs. 1 and 2 to take into account possible photometric discrepancies
due to the use of different telescopes and instruments.

The $R$ band light curve, the best sampled, exhibits in its early portion
a flaring activity with hour time scale (Fig. 1), with an initial increase
(confirmed by S.G. Bhargavi, priv. comm.). 
The flux then shows a slow decline, lasting about 1.5 days and
following approximately a power-law, with a slope $\alpha \sim$ 0.7 ($f(t)
\propto (t-t_0)^{-\alpha}$, where $t_0$ is the GRB trigger time).
Subsequently the light curve flattens, and an approximately constant, or
slightly increasing, behavior is seen till around 3.7 days after the GRB.
%After this light local maximum, the flux starts decreasing again.  
The flux starts decreasing again thereafter. This decline, which can be
fitted by a very steep power-law (index $\alpha \sim$ 3.5), levels off
around 5 days after the GRB trigger to a ``plateau" of two days duration.
The flux resumes then the decreasing
trend, with a shallower power-law of $\alpha \sim$ 3, till the end of the
monitoring. 
The late $R$ band epoch flux and upper limit reported by
Fruchter et al. (2000) and Veillet (2000), respectively, are consistent
with this trend.

The $B$ band light curve appears well correlated with the $R$ band, though
less well sampled. The $B$ band points at 3.5 and 3.7 days after the
GRB suggest a variation opposite to that observed simultaneously in the
$R$ band. However, the $B$ band variation is not significant and
determines only a marginally significant change
in the $B-R$ color (Fig. 2a).  The $B$ band upper limit determined
on April 5 with the TNG is consistent with the power-law decline of the
final portion of the light curve.

The fewer $V$ band points show a good correlation of the light curve
with that in the $R$ band, with no measurable temporal lag (the $V-R$
color is unchanged, Fig. 2b). 
In particular, the $V$ band data around 3-4 days after the high-energy
event also suggest a local flattening of the light curve. 

The $I$ band data confirm the general steepening observed in the other
bands, although the second plateau at 5-7 days after the high-energy event
is less clearly seen than in the $B$ and $R$ light curves.
Also, a rapid flux increase is apparent at the beginning of the $I$ band
monitoring, delayed by $\sim$7 hours with respect to that seen in the
$R$ band light curve (see Figs. 1 and 2c).

\section{Discussion}

Our optical monitoring of the bright GRB000301C afterglow has provided
one of the best sampled afterglow datasets, especially in the $R$ filter. 
The long term behavior of this optical afterglow is better
described by a continuous steepening, rather than by a single power-law,
as expected in afterglows developing in laterally spreading jets (Sari et
al. 1999; Rhoads 1999) or decelerating to non-relativistic regimes (Dai \&
Lu 1999), and seen in few other cases.

Among equally well monitored GRB afterglows, GR000301C appears peculiar in
that several shorter time scale variations are superimposed on the long
term decrease.  The reality of two of these (3.1-3.7 days and 5-7 days
after the event) is supported by their appearance in more than one band.
The first two points of the $R$ and $I$ band light curves might suggest a
rise and could be reminiscent of the early (1-2 days after the GRB
trigger) light curve of GRB970228 and GRB970508 (Guarnieri et al. 1997;
Pedersen et al. 1998), although in the latter the initial increase was
more structured. In the present case we cannot
exclude that the flux is declining since the start of the monitoring, and
hour time scale flares modulate this decrease.  Some isolated short term
variability events are seen in GRB980703 (Vreeswijk et al. 1999) and
GRB990123 (Castro-Tirado et al. 1999) and are almost totally absent in
GRB990510 (e.g., Stanek et al. 1999).  

Recently, various scenarios have been developed in which intrinsic
re-energization of the blast wave, or irregularities of the dense
interstellar medium in which the blast is expanding can account for the
observed behavior (Panaitescu et al. 1998;  M\'esz\'aros et al. 1998; 
Sari \& M\'esz\'aros 2000; Wang \& Loeb 2000; Dai \& Lu 2000).  In
particular, a flattening of the afterglow light
curve, similar to that exhibited by GRB000301C in the $R$ band on days
3.1-3.7 and 5-7 days after the GRB, is predicted by Kumar \& Piran (2000)
as a consequence of the collision of a slow shell ejected at a late time
after the GRB with an outer shell decelerated by its propagation in the 
circumburst medium (see their Fig. 5). We note that the temporal
occurrence of the observed flattenings could be consistent with a
``colliding shells" interpretation, while it is incompatible with the time
scale implied by an hypernova scenario (see Rhoads \& Fruchter 2000). 

The lack of a clear correlation between the $R$ band light curve and the
$I$ and $K$ band light curves (see Rhoads \& Fruchter 2000 for the latter) 
might be due to non strict simultaneity of the data points.  In fact, the
$R-I$ and $R-K$ colors as a function of time show only marginally
significant deviations from constancy (Fig. 2c and 2d), and these are
mainly exhibited by color values derived from pairs of measurements
separated in time by more than 0.5 hours, the shortest variability time
scale observed in this afterglow. Our findings underline the critical
importance of intensive multiwavelength observations of afterglow sources.

\begin{acknowledgements}

We thank the staff of the Loiano, Calar Alto, Sierra Nevada, UPSO and TNG
Observatories. CB, AG, and AP acknowledge the University of Bologna
(Funds for Selected Research Topics). GMB thanks the Russian Fund of
Fundamental Researches for support (grant 98-02-17570). 

\end{acknowledgements}


\begin{thebibliography}{}

\bibitem{} Berger E., Frail D.A., 2000, GCN 589

\bibitem{} Bernabei S., Marinoni S., Bartolini C., et al. 2000a, GCN 571

\bibitem{} Bernabei S., Bartolini C., Di Fabrizio L., et al. 2000b, GCN
 599

\bibitem{} Bertoldi F., 2000, GCN 580

\bibitem{} Castro S.M., Diercks A., Djorgovski S.G.et al., 2000, GCN 605

\bibitem{} Castro-Tirado A.J. et al., 1999, Science 283, 2069

\bibitem{} Dai Z.G., Lu T., 1999, ApJ 519, L155

\bibitem{} Dai Z.G., Lu T., 2000, ApJ submitted (astro-ph/0005417)

\bibitem{} Fruchter A.S. et al., 2000, GCN 701

\bibitem{} Fynbo J.P.U. et al., 2000, GCN 570

\bibitem{} Guarnieri A., Bartolini C., Masetti N. et al., 1997, A\&A 328,
L13

\bibitem{} Henden A., 2000, GCN 583

\bibitem{} Jensen B.L. et al., 2000, A\&A, submitted (astro-ph/0005609)

\bibitem{} Kumar P., Piran T., 2000, ApJ 532, 286

\bibitem{} M\'esz\'aros P., Rees M.J., Wijers, R.A.M.J., 1998, ApJ 499,
	301

\bibitem{} Panaitescu A., M\'esz\'aros P., Rees M.J., 1998, ApJ 503, 314

\bibitem{} Pedersen H., Jaunsen A.O., Grav T. et al., 1998, ApJ 496, 311

\bibitem{} Rhoads J.E., Fruchter A.S., 2000, ApJ, submitted 
	(astro-ph/0004057)

\bibitem{} Rhoads J.E., 1999, ApJ 525, 737

\bibitem{} Sagar R. et al., 2000, Bull. Astr. Soc. India, submitted
	(astro-ph/0004223)

\bibitem{} Sari R., Piran T., Halpern J.P., 1999, ApJ 519, L17

\bibitem{} Sari R., M\'esz\'aros P., 2000, ApJ 535, L33

\bibitem{} Schlegel D.J., Finkbeiner D.P., Davis M., 1998, ApJ 500, 525

\bibitem{} Smette A., Fruchter A.S., Gull T. et al., 2000, GCN 603

\bibitem{} Smith D.A., Hurley K., Cline T., 2000, GCN 568

\bibitem{} Stanek K.Z., Garnavich P.M., Kaluzny J. et al., 1999, ApJ 522,
L39

\bibitem{} Stetson P.B., 1987, PASP 99, 191

\bibitem{} Veillet C., 2000, GCN 623

\bibitem{} Vreeswijk P.M., Galama T.J., Owens A.O., et al., 1999, ApJ
523, 171

\bibitem{} Wang, X., \& Loeb, A. 2000, ApJ 535, 788

\end{thebibliography}
\end{document}